\documentclass[conference]{IEEEtran}
\IEEEoverridecommandlockouts
\usepackage{cite}
\usepackage{amsmath,amssymb,amsfonts}
\usepackage{algorithmic}
\usepackage{graphicx}
\usepackage{courier}
\usepackage{pifont}
\usepackage{censor}
\usepackage{booktabs}
\usepackage{hyperref}
\usepackage{multirow}
\usepackage{textcomp}
\usepackage{url}
\usepackage[table]{xcolor}
\def\BibTeX{{\rm B\kern-.05em{\sc i\kern-.025em b}\kern-.08em
    T\kern-.1667em\lower.7ex\hbox{E}\kern-.125emX}}
\begin{document}

\definecolor{strong_purple}{RGB}{97, 75, 195}
\definecolor{strong_green}{RGB}{51, 187, 197}
\definecolor{medium_green}{RGB}{133, 230, 197}
\definecolor{light_green}{RGB}{200, 255, 224}


\makeatletter
    \newcommand{\linebreakand}{%
      \end{@IEEEauthorhalign}
      \hfill\mbox{}\par
      \mbox{}\hfill\begin{@IEEEauthorhalign}
    }
    \makeatother

\title{DID Link: Authentication in TLS with Decentralized Identifiers and Verifiable Credentials\\
}

\author{\IEEEauthorblockN{Sandro Rodriguez Garzon}
\IEEEauthorblockA{\textit{Service-centric Networking} \\
\textit{Technische Universität Berlin/T-Labs}\\
Berlin, Germany \\
sandro.rodriguezgarzon@tu-berlin.de}
\and 
\IEEEauthorblockN{Dennis Natusch}
\IEEEauthorblockA{\textit{Service-centric Networking} \\
\textit{Technische Universität Berlin/T-Labs}\\
Berlin, Germany \\
d.natusch@campus.tu-berlin.de}
\and
\IEEEauthorblockN{Artur Philipp}
\IEEEauthorblockA{\textit{Service-centric Networking} \\
\textit{Technische Universität Berlin/T-Labs}\\
Berlin, Germany \\
a.philipp@tu-berlin.de}
\and
}

\author{\IEEEauthorblockN{Sandro Rodriguez Garzon, Dennis Natusch, Artur Philipp, Axel Küpper}
\IEEEauthorblockA{\textit{Service-centric Networking} \\
\textit{Technische Universität Berlin / Telekom Innovation Laboratories (T-Labs)}\\
Berlin, Germany \\
\{sandro.rodriguezgarzon\}\textbar\{a.philipp\}\textbar\{axel.kuepper\}@tu-berlin.de, d.natusch@campus.tu-berlin.de}
\and 
}

\author{\IEEEauthorblockN{Sandro Rodriguez Garzon}
\IEEEauthorblockA{\textit{Service-centric Networking} \\
\textit{Technische Universität Berlin/T-Labs}\\
Berlin, Germany \\
sandro.rodriguezgarzon@tu-berlin.de}
\and 
\IEEEauthorblockN{\hspace{0.3cm}Dennis Natusch}
\IEEEauthorblockA{\hspace{0.3cm}\textit{Service-centric Networking} \\
\hspace{0.3cm}\textit{Technische Universität Berlin}\\
\hspace{0.3cm}Berlin, Germany \\
\hspace{0.3cm}d.natusch@campus.tu-berlin.de}
\and
\IEEEauthorblockN{Artur Philipp}
\IEEEauthorblockA{\textit{Service-centric Networking} \\
\textit{Technische Universität Berlin/T-Labs}\\
Berlin, Germany \\
a.philipp@tu-berlin.de}
\and
\linebreakand 
\IEEEauthorblockN{\hspace{-0.9cm}Axel Küpper}
\IEEEauthorblockA{\hspace{-0.9cm}\textit{Service-centric Networking} \\
\hspace{-0.9cm}\textit{Technische Universität Berlin/T-Labs}\\
\hspace{-0.9cm}Berlin, Germany \\
\hspace{-0.9cm}axel.kuepper@tu-berlin.de}
\and
\IEEEauthorblockN{\hspace{-0.1cm}Hans Joachim Einsiedler}
\IEEEauthorblockA{\textit{Group Technology} \\
\textit{Deutsche Telekom AG}\\
Berlin, Germany \\
hans.einsiedler@telekom.de}
\and 
\IEEEauthorblockN{Daniela Schneider}
\IEEEauthorblockA{\textit{Core \& Network Services} \\
\textit{Deutsche Telekom AG}\\
Vienna, Austria \\
daniela.schneider@magenta.at}}

\newcommand{\ts}{\textsuperscript}

\maketitle

\begin{abstract}

Authentication in TLS is predominately carried out with X.509 digital certificates issued by certificate authorities (CA). The centralized nature of current public key infrastructures, however, comes along with severe risks, such as single points of failure and susceptibility to cyber-attacks, potentially undermining the security and trustworthiness of the entire system. With Decentralized Identifiers (DID) alongside distributed ledger technology, it becomes technically feasible to prove ownership of a unique identifier without requiring an attestation of the proof's public key by a centralized and therefore vulnerable CA. This article presents DID Link, a novel authentication scheme for TLS 1.3 that empowers entities to authenticate in a TLS-compliant way with self-issued X.509 certificates that are equipped with ledger-anchored DIDs instead of CA-issued identifiers. It facilitates the exchange of tamper-proof and 3rd-party attested claims in the form of DID-bound Verifiable Credentials after the TLS handshake to complete the authentication with a full identification of the communication partner. A prototypical implementation shows comparable TLS handshake durations of DID Link if verification material is cached and reasonable prolongations if it is obtained from a ledger. The significant speed improvement of the resulting TLS channel over a widely used, DID-based alternative transport protocol on the application layer demonstrates the potential of DID Link to become a viable solution for the establishment of secure and trustful end-to-end communication links with decentrally managed digital identities.

\end{abstract}


\section{Introduction}

Today, the Transport Layer Security (TLS) protocol \cite{IETF.tls} is widely used in conjunction with TCP or QUIC to ensure secure end-to-end communication between two parties, commonly referred to as client and server. During the TLS handshake, a shared secret is established among the involved parties, enabling the encryption and decryption of the information exchanged during the session, ensuring confidentiality and integrity of the data transmitted. In addition, TLS prescribes the server to authenticate itself during the TLS handshake preferably with a digital certificate according to the X.509 standard \cite{IETF.x509}. Mutual TLS (mTLS) refers then to a variant of TLS where both parties authenticate with digital certificates. 


An X.509 digital certificate binds an identifier of an entity to a public key and is digitally signed by a certificate authority (CA). The key enables the entity owning the associated private key to prove during the TLS handshake that it is indeed the subject referred to by the identifier. The digital signature serves as an approval of the binding by a trustful 3rd party and ensures the integrity and authenticity of the certificate. The value of a digital certificate for trust establishment purposes, however, is inherently related to the trustworthiness of the CA \cite{specter2016economics}. A CA is a pivotal element of a public key infrastructure (PKI), responsible to issue, manage, and revoke digital certificates. Today's PKIs for the Web are governed and operated centrally by a few organizations such as Let's Encrypt, DigiCert, and GlobalSign \cite{Aqsa:2020}. This centralization of trust creates potential single points of failures and vulnerabilities, where the compromise or failure of a CA or the whole PKI undermines the security and trustworthiness of the entire system \cite{Delignat.2014, Amann:2013}, making centralized PKIs perfect targets for cyberattacks and governmental interventions. Breaches in centralized PKIs occur from time to time and cause an immense damage to the digital economy and society as a whole \cite{2013.Meulen, us.2024}.

With Decentralized Identifiers (DID) \cite{WorldWideWebConsortium.822021}, the W3C provides an important building block to realize the vision of a decentralized PKI. The core idea is that each entity remains in full control of its own identifier, denoted as the DID, and related verification material. No party other than the entity itself is involved in the creation and management of its DID and the verification material used for the ownership proof. The uniqueness of the identifier and the binding to the verification material is made verifiable for others by sharing both artifacts via a decentralized verifiable data registry (VDR) that is commonly governed and operated by multiple parties. The decentralized VDR ensures algorithmically that everyone is able to access the verification material associated with any DID but no one besides the owner of a DID is capable to alter this binding in the VDR. Since DIDs are not issued by CAs, impersonation attacks become only technically feasible if either the DID owner's secure vault with the private key or the decentrally operated VDR as a whole is compromised.

Authentication is the process of revealing the digital identity and proving to be the subject the digital identity represents in the digital realm. However, neither the DIDs nor the verification material in the VDR reveal the digital identity of the owner. A DID owner can prove DID ownership but has no means to trustfully share its claims about its digital identity with others. A W3C recommended verifiable credential (VC)\cite{WorldWideWebConsortium.05.11.2021} fills this gap and is a secure container, comprising one or multiple attested identity attributes of an entity's digital identity. It is cryptographically linked to the entity's DID and digitally signed and issued by a reputable 3rd party. In conjunction with the DID ownership proof, it can be used by the DID and VC owner for authentication or authorization purposes. VCs serve a similar purpose as Attribute Certificates as defined in RFC 5755\cite{IETF.authorization}, but they are flexible with respect to their structure, content, and intended usage.

Unfortunately, TLS in its latest incarnation does not support the use of DIDs and VCs as an alternative to X.509 certificates or raw public keys \cite{IETF.raw} for the authentication. The rather strict specifications offer only limited to no room for groundbreaking changes that would be required to handle artifacts of decentrally managed digital identities within the authentication scheme of TLS. Nevertheless, the use of X.509 certificates for authentication purposes in TLS will not cease any time soon due to their widespread adoption, mature standardization, and established PKIs. So as long as TLS does not natively support decentralized identity concepts, an interim solution for the authentication of entities with DIDs and the identification with VCs on the transport layer is required. Otherwise, the decentralization efforts taken on the application layer with the introduction of distributed ledger technology and dApps in the context of Web 3.0 are hampered by restrictions of the build-in security of today's predominant transport protocols.

This article presents DID Link, a novel authentication scheme for TLS that empowers both parties to alternatively authenticate either with pseudo-anonymous DIDs or in combination with VCs instead of using CA-issued X.509 certificates or raw public keys. With DID Link, self-issued X.509 certificates are used by the involved parties as a container to communicate their DIDs and associated verification material without violating the X.509v3 specification. This makes it possible with DID Link to rely on the established way parties are authenticated during the TLS handshake by only - in addition - integrating a custom certificate validation process tailored to DID-equipped X.509 certificates. Atop the pseudo-anonymous authentication during the TLS handshake with DIDs, in DID Link, the parties can additionally exchange DID-bound VCs afterwards, to complete the authentication scheme with a clear identification of subjects that are referred to by the pseudo-anonymous DIDs used during the TLS handshake. Since DID Link remains fully compliant with TLS 1.3, the parties establishing a TLS connection with DID Link benefit from the widespread support, level of standardization, and thoroughly investigated security of TLS with X.509-based authentication and, at the same time, take advantage of the key properties of decentralized identity management, namely increased data privacy and protection, improved robustness against failures, and enhanced security due to a lack of a single point of vulnerability.

\section{TLS handshake}
\label{sec:handshake}



TLS is a security protocol situated at the transport layer, aiming to facilitate a secure end-to-end communication in an untrusted network. It encrypts the data being exchanged, preserving confidentiality and integrity of the communication link between two parties, typically referred to as the client and the server. During the TLS handshake, both parties negotiate the security parameters of the session, authenticate the server (and optionally the client), and establish a shared master secret that is used to derive keys for the encryption and data integrity checks. This process involves a series of clearly defined handshake protocol messages exchanged in a specified temporal order between both parties. The (m)TLS 1.3 handshake with certificate-based authentication is illustrated in Fig. \ref{fig:tls-handshake}.

\begin{figure}[t!]
\centerline{\includegraphics[scale=1]{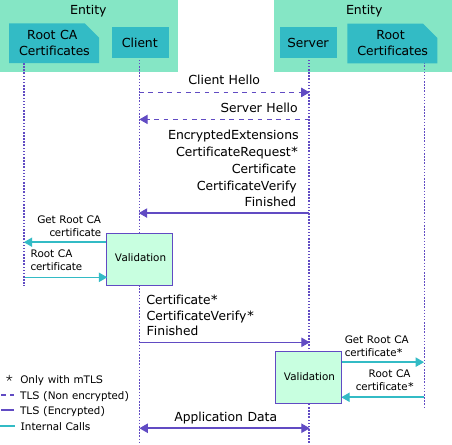}}
\caption{(m)TLS 1.3 handshake}
\label{fig:tls-handshake}
\end{figure}


The ClientHello message, which is the first message sent by the client to initiate the TLS handshake, contains, beside other information, the proposed security parameters for the Diffie–Hellman key exchange (DHE). As a response, the ServerHello message comprises, beside other information, the selected security parameters for the DHE. At this point, the server can already derive the master secret and therefore encrypts all subsequent handshake messages. Additional data that is not required for the key agreement purposes can be sent encrypted with the EncryptedExtensions message. If no additional data is needed, the message is empty. Upon receipt of the ServerHello message, the client derives the master secret based on the server-selected security parameters and thus becomes able to encrypt and decrypt all the following ciphered handshake messages. 

With the optional CertificateRequest message, the server requests a certificate from the client for authentication purposes (with mTLS). The mandatory Certificate message from the server contains the X.509 certificate of the server and potentially certificates of intermediate CAs while the CertificateVerify message contains a cryptographic proof that the server is in possession of the private key corresponding to the certificate's public key. This proof in form of a digital signature also serves as an integrity and authenticity proof over the handshake messages exchanged up to this point. The Finished message contains a similar proof and marks the completion of the TLS handshake from the server's perspective. 

Next, the client runs through a couple of validation steps, collectively referred to as validation in Fig. \ref{fig:tls-handshake}. This encompasses the validation of the certificate, the CertificateVerify, and the Finished message. With the validation of the certificate it is ensured that the certificate's integrity is given, that it is not expired or revoked by the CA, and that it is trustworthy. It comprises of a check of the digital signature, the certificate's validity period, and the revocation status as well as a verification of the chain of certificates from intermediary CAs, all the way up to the root CA. The trustworthiness is then given if the root CA as a common trust anchor is declared as trustworthy by the entity itself. However, at this point, the server has not been authenticated yet. It is only with the affirmative verification of the CertificateVerify message that the server's identity is conclusively authenticated to the client during the TLS handshake. If no client authentication is requested by the server, then the client will complete the TLS handshake with a Finished message. Otherwise, the client compiles and sends a Certificate message, containing the client's X.509 certificate, and a CertificateVerify message prior to completing the TLS handshake with the Finished message to the server. The server will conduct the validation of the message(s) and a potential certificate provided by the client in an analog manner.    

\section{Decentralized Identifiers and Verifiable Credentials}

With Decentralized Identifiers (DID)\cite{WorldWideWebConsortium.822021} and Verifiable Credentials (VC) \cite{WorldWideWebConsortium.05.11.2021}, the W3C has laid the foundation for the decentralization of identity management in the future Web. A DID is a unique alphanumeric string of the form \texttt{did:[method]:[subject]} that refers to a DID subject, e.g., a human or a thing, and resolves to a single unique DID document (abbr. as DID doc). The DID method specifies the mechanism for creating, resolving, and updating the DID doc while the subject-specific part, directly following the DID method, uniquely identifies the subject itself. The document the DID resolves to contains mainly verification material such as multiple public keys of the subject the DID refers to. It empowers the subject in possession of the corresponding
private keys to cryptographically prove that it is indeed the DID owner. To conduct this proof, however, the DID doc must be accessible by a verifying party and it needs to be guaranteed that the resolved DID doc is unique, up-to-date, and not tampered with by others than the DID owner itself. This becomes possible through a VDR. A VDR persists DID docs similar to a database and makes them available to others. It guarantees that there exists only a single DID doc for each DID and that only the DID owner or a deputy selected by the DID owner is able to modify it while all others can read it. Decentralized VDRs that are commonly operated and governed by multiple parties can be realized as decentralized file systems or distributed ledgers. Since DID docs in a VDR are accessible by everyone with access to the VDR, it is important for privacy and data protection purposes to consider the DID as a pseudo-anonymous identifier. The associated DID doc shall therefore not contain any information that reveals the identity being referred to by the DID. With DIDs, a subject's identifier becomes owned and fully controlled by the subject itself in order to break free from the dependency on centralized CAs to issue identifiers. Self-signed X.509 certificates are the closest analog to DIDs but they do not guarantee out-of-the-box uniqueness of the identifier because two self-signed X.509 certificates from different parties might contain the same identifier in the subject field. It is also not possible to encode multiple public keys for different purposes in a self-signed X.509 certificate, as is possible with DID docs.

\begin{figure}[t!]
\centerline{\includegraphics[scale=0.9]{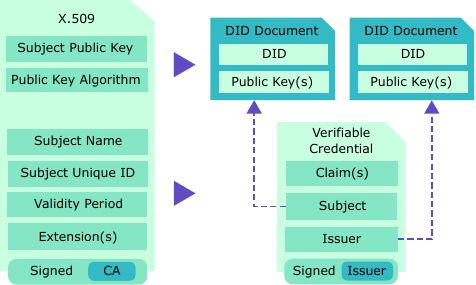}}
\caption{X.509 digital certificates and their conceptual relation to Decentralized Identifiers and Verifiable Credentials}
\label{fig:relation}
\end{figure}

A DID and the associated DID doc enable the DID owner to prove ownership of the DID. Yet, a DID is pseudo-anonymous and gives the verifying party no clue about the subject owning the DID besides the fact that the subject owns the DID. A verifiable credential (VC) is a way for a DID owner to bilaterally share verifiable claims about its identity with others. It materializes as a digital document that contains the subject's identifier in form of the subject's DID and one or more claims about the identity. It is not self-issued by the subject, but by an issuer who verifies the claims about the subject and confirms their integrity and authenticity by generating a proof, such as a digital signature, which becomes part of the VC. Fig. \ref{fig:relation} illustrates the conceptual relation between an X.509 certificate and the DIDs and VCs. Attribute certificates, as specified in RFC 5755 \cite{IETF.authorization}, serve a specialized function akin to VCs. However, these certificates are not as widely recognized or utilized. Unlike VCs, which boast a broader scope of flexibility and are not confined to a singular purpose, attribute certificates are tailored specifically for authorization.


During the issuance process, a VC gets created and transferred from the issuer to the subject (which becomes the holder of the VC), granting it autonomy to decide about the storage location and to select individuals or entities with whom to share it for whatever purpose. An exemplary protocol for the credential issuance process between the issuer and the holder on the application layer is the Issue Credential Protocol 2.0\footnote{\url{https://github.com/hyperledger/aries-rfcs/blob/main/features/0453-issue-credential-v2/README.md}} of the Hyperledger Aries project. Once an entity receives a VC within a Verifiable Presentation (VP) from a VC holder, possibly for authentication purposes, it can ascertain the authenticity and integrity of the shared claims by means of the verification material found in the issuer's DID doc. Hence, an entity does not need to be in possession of a DID in order to act in the role of a verifier for VCs but it needs at least access to the DID docs in a VDR. Potential application layer presentation protocols are the DIF Presentation Exchange\footnote{\url{https://identity.foundation/presentation-exchange/}} or Present Proof Protocol 2.0\footnote{\url{https://github.com/hyperledger/aries-rfcs/tree/main/features/0454-present-proof-v2}} of the Hyperledger Aries project. Figure \ref{fig:ssi} illustrates the issuance of VCs from the issuer to the holder and the presentation of VCs from the holder to the verifier. During the presentation, VPs can contain multiple VCs from the same and/or different issuers. The Verifiable Credentials Model of the W3C only describes the required ingredients and general structure of a VC, giving developers the freedom to select a technical format of choice depending on the requirements. For example, VCs in form of SD-JWTs\cite{IETF.sdJWTs} support the selective disclosure of claims in VPs while JSON-LD\cite{WWW.JSON-LD} puts the encoded claims into context.

\begin{figure}[t!]
\centerline{\includegraphics[scale=0.9]{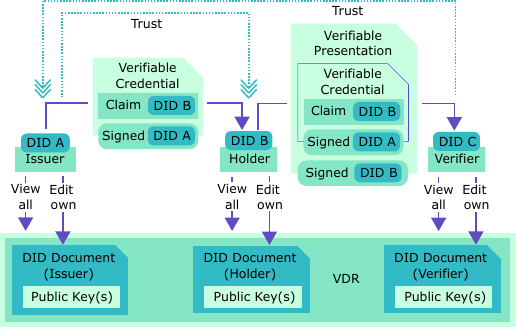}}
\caption{Bilateral off-ledger issuance and presentation of Verifiable Credentials that are cryptographically bound to VDR-anchored Decentralized Identifiers.}
\label{fig:ssi}
\end{figure}



\section{Related Work}
First attempts to make use of an early predecessor of DIDs for authentication purposes in TLS 1.2 where discussed at the Internet Identity Workshop in 2017, resulting in a first rudimentary draft of a DID TLS Specification\cite{DID-TLS.2017}. It sketches the idea of deriving X.509 certificates presented during the TLS handshake from the verification material of a DID doc that is anchored in a distributed ledger. The server name indication (SNI) extension is used in the ClientHello message to signalize the DID the server has to use for the certificate derivation and authentication process. The extension gets encrypted with verification material of the server as found in the server's DID doc. Beckwith et al. introduces a new type of certificate for resource-constraint IoT devices in TLS that comprises of the device's public key, the device owner's public key, the key exchange algorithm and parameters for the algorithm \cite{Beckwith.2020}. The validation of the ownership presented in the certificate and the authentication of the device during the modified TLS handshake is conducted with the device ownership information that are anchored in a distributed ledger by the device's owner. In \cite{Chu.2020}, Chu et al. uses self-signed X.509 certificates for authentication purposes in TLS 1.2. The issuer and subject fields, however, contain an identifier that is compliant with the X.500 naming scheme and is anchored in a distributed ledger together with public key material used to create and sign the certificate. During certificate validation, the latest transaction containing the identifier and the public keys is queried in the distributed ledger and its elements being compared with the self-signed certificate. Yan et al. describes an approach in which the self-signed certificate of a client is anchored in a distributed ledger as a whole \cite{Yan.2020}. In a modified TLS 1.2 handshake, the identifier of the certificate is then shared by the client within the ClientHello message, giving the server the opportunity to grab the whole certificate and its status from the ledger to proceed with the authentication of the client in compliance with mTLS 1.2. The modified access control mechanism for the Modbus IIoT protocol presented in \cite{Lorenzo.2021} uses X.509 certificates for authentication in mTLS that include the DIDs of the client and server in the Subject Alternative Name (SAN) extensions of the certificates. The DID doc are anchored in a VDR and queried for the certificate validation during the TLS handshake. However, it is not described how the identity validation with the help of the VDR is realized. 

A brief discussion about the potential of combined usages of X.509 and DID/VCs is given by Bastian et al. in \cite{Bastian.2022}. Although not explicitly targeting the usage of DID-enhanced X.509 certificates for authentication purposes in TLS, it is described, among other options, how DIDs can be derived from X.509 key pairs and how signed DIDs can be embedded into X.509 certificates. Claudio et al. argues that the TLS handshake with TLS 1.3 and X.509 certificates has to be modified so that DID documents can be resolved by the software artifact implementing TLS \cite{Alessio.2023}. As a proof of concept, they realized a DID resolver within the OpenSSL implementation of TLS 1.3. But due to the proliferation of DID methods and sometimes application-specific constraints with respect to a DID method, it remains questionable whether the potential speed improvement by implementing the DID resolution at the transport layer outweighs the flexibility of conducting it on the application layer. In \cite{PERUGINI.2023}, Perugini et al. introduce a modified TLS 1.3 handshake that enables client and server to mutually authenticate with X.509, DIDs and/or VCs. A mixed operation with the client using an X.509 certificate and the server using a DID and a VC for authentication and vice versa is supported as well. With the DID authentication mode, the server and optionally the client send a custom DID message instead of an X.509 certificate in the Certificate message, containing the DID method and the DID itself. DIDs are hereby assumed to be anchored in a distributed ledger and resolved in the TLS handshake during the certificate validation step. Despite being the first promising attempt to fully integrate DIDs and VCs into the TLS handshake for authentication purposes, it requires the TLS handshake to be significantly modified, loosing compliance with TLS 1.3. 

\section{Concept}

With the aforementioned approaches \cite{DID-TLS.2017, Beckwith.2020, Chu.2020, Yan.2020, Alessio.2023}, DIDs, their early predecessors, or similar concepts were used to conduct a pseudo-anonymous authentication within a TLS handshake, considering a distributed ledger as the root of trust for verification material. The process to identify the subject the DID refers to, however, was not addressed nor discussed. Only Perugini et al. \cite{PERUGINI.2023} consider the identification as an essential component of the whole authentication process. They make use of VCs within the TLS 1.3 handshake to complete the authentication with the identification of the subject. However, VCs contain highly application-specific content, are of unpredictable size, and can materialize themselves in various formats \cite{young.2021}. The identification of a subject with VCs should therefore be conducted preferably immediately after the TLS handshake. Ideally, using TLS records for the sole purpose of carrying messages related to the identification process. The introduction of a new type of TLS record comes with the advantage that all messages related to the identification process are semantically separated from any application data, introducing a new identification sub layer in TLS. Fortunately, the specification of TLS 1.3 permits the introduction of new types of TLS records as long as its existence and usage is agreed upon within the TLS handshake. 



With DID Link, each entity - the client and the server - is assumed to be equipped with its own DID. The corresponding DID docs are anchored in a commonly operated and governed distributed ledger, accessible by all entities. The pseudo-anonymous authentication of an entity occurs within the TLS 1.3 handshake. Subsequently, the identification of the subject takes place at the new identification sub layer via the exchange of VCs immediately following the successful completion of the TLS handshake. The identification step is optional because both parties might know each other and each others' DIDs from past interactions. Similar to the approach presented in \cite{Chu.2020}, self-signed X.509 certificates are used in DID Link during the TLS handshake as a standardized means to communicate the identifier as well as the public key of the certificate's subject. In contrast to \cite{Chu.2020}, a W3C DID is used as an identifier. It is stored in the SAN extension instead of the subject field, as sketched in \cite{Lorenzo.2021}, due to the strict naming rules of the X.509 subject field and the 64 byte limit of the common name (CN) attribute. The public key of the self-signed X.509 certificate corresponds to a public key of the subject's DID doc in the distributed ledger. An exemplary message exchange of DID Link, comprising the pseudo-anonymous authentication with DIDs in the TLS handshake and the identification with VCs at the application layer, is illustrated in Fig. \ref{fig:concept}. 

\begin{figure}[t!]
\centerline{\includegraphics[scale=1.1]{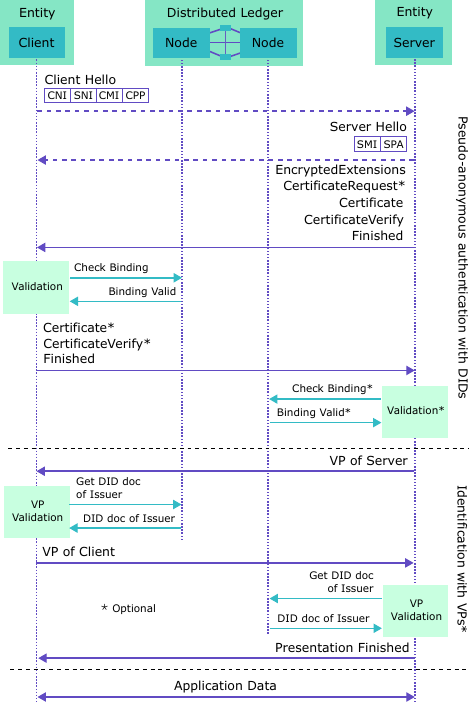}}
\caption{Using DIDs in X.509 certificates during the TLS handshake for the pseudo-anonymous authentication and VPs afterwards to (mutually) identify entities after the TCP connection establishment with DID Link.}
\label{fig:concept}
\end{figure}

As described in Section \ref{sec:handshake}, the client and server verify each others certificate, the CertificateVerify and the Finished messages. With a CA-issued X.509 certificate, the integrity needs to be tested by verifying the digital signature of the CA. Moreover, if intermediary CAs are present then the chain of certificates needs to be verified all way up to the root CA. However, if a self-signed X.509 certificate's subject refers to a DID, the certificate validation becomes significantly simpler. The integrity of a self-signed and DID-based X.509 certificate is a prerequisite, but does not get explicitly tested through the verification of the certificate's digital signature. DID Link uses the certificate only as a vehicle to communicate the binding. However, the validity of the binding between the given DID in the SAN field and the public key in the Subject Public Key Info field needs to be checked with other means. Since the distributed ledger guarantees the uniqueness of a DID and that only the DID owner is able to alter the corresponding DID doc, it can, in the role of an independent technical trust anchor, confirm or deny the validity of a binding between a DID and a public key. If a binding's validity is confirmed by or through the distributed ledger, the verifying entity still does not know whether the other entity is the DID owner or an entity just pretending to own the DID. The DID ownership proof is missing. Since the CertificateVerify message is signed with the private key corresponding to the public key of the certificate, it serves as a valid DID ownership proof. Technically. the self-signed X.509 certificate would actually only need to contain the DID, since the public key could be queried by resolving the DID doc from the ledger. However, providing the public key up front within the certificate speeds up the validation process because the verification of the CertificateVerify message can be carried out in parallel to the binding check via the ledger. The order of validation checks is therefore irrelevant. If the binding and the CertifyVerify message are both valid, the pseudo-anonymous authentication with a DID is completed. 

Typically, the client is already aware of the server's DID and latest DID doc. It can therefore decide to skip the binding check during the TLS handshake. The same applies to the binding check at the server if the client's DID and latest DID doc are known by the server. If a cached DID doc of the server at the client is outdated according to local security policies, the latest version can also be resolved before even starting the TLS handshake. If the server owns multiple DIDs and if the client expects a particular DID of the server to be used for the pseudo-anonymous authentication, then the client can make use of the \textit{Server Name Indication (SNI)} extension in compliance with RFC 6066 \cite{IETF.extensions} to indicate the desired server DID in the ClientHello message, similar to \cite{DID-TLS.2017} but not encrypted. Analogously, the client can optionally communicate its own DID in the ClientHello message within a new extension denoted as \textit{Client Name Indication (CNI)}. But it needs to be noted that the communication of the source and target DIDs in the non-encrypted ClientHello message enables eavesdroppers to trace interactions among entities.    

As of this moment, there exist a massive number of DID methods \cite{Hoops.2023, Bistarelli.2023} with each one implementing the CRUD operations for the resolution of a DID doc in a particular way, e.g., using distributed ledger technology, a distributed hash table, or web services as the VDR. With the new extension \textit{Client DID Methods Indication (CMI)} in the ClientHello message, the client becomes able to share a list of DID methods that the client supports. The server then responds with the \textit{Server DID methods Indication (SMI)} extension within the ServerHello message, containing a subset of the DID methods listed in the CMI extension that the server supports. Similar extensions were introduced in \cite{PERUGINI.2023} to give the server early on the chance to abort the TLS handshake in case there exists no overlap amongst the DID methods the entities support. In contrast to the solution presented by \cite{PERUGINI.2023}, the extensions also address the accessible VDRs because both entities might support the same set of DID methods but the accessibility to a VDR might not be given on both sides.        

Once the TLS handshake is done, VPs can optionally be exchanged for identification purposes. The type, order, and format of messages for the presentation of VPs is defined by a presentation protocol. Most presentation protocols are agnostic to the VP's format. As outlined above, there exist multiple presentation protocols and VP formats, with different sets of features for different use cases. To maintain flexibility in the choice of the presentation protocol and VP format, but still make the exchange of VPs on the new identification sub layer predictable for the entities, the latter can agree on a VP protocol during the TLS handshake. The Application-Layer Protocol Negotiation (ALPN) Extension according to RFC 7301 is designed for the negotiation of the application layer protocol in TLS. However, the identification is semantically located at a sub layer within TLS. So instead of making use of ALPN to agree on a presentation protocol and to potentially break legacy systems that make use of ALPN for the purpose it was meant to be used, a new extension was introduced for the ClientHello and ServerHello messages. With the new extension, the client proposes a list of supported presentation protocols within the \textit{Client Presentation Proposal (CPP)} extension and the server confirms the one agreed upon within the ServerHello message in the \textit{Server Presentation Agreement (SPA)}. Both extensions are mandatory in case an identification is about to happen after the TLS handshake because their mere existence implicitly legitimizes the use of the new identification sub layer. Although the presentation of VPs in Fig. \ref{fig:concept} happen sequentially, with the server presenting its VP first, the order of presentations is not prescribed. The client might present first or both presentations can happen in parallel, to speed up the identification process.   

\begin{table}[h]
\vspace{0.2cm}
\centering
\begin{tabular}{|c|c|c|c|c|c|c|}
\hline
\rowcolor{strong_green} \multicolumn{2}{|c|}{\textbf{Authentication}} & \multicolumn{4}{|c|}{\textbf{Extensions}} & \cellcolor{white}\\ 

\hline

\rowcolor{medium_green} \textbf{Client} & \textbf{Server} & \textbf{SNI} & \textbf{CNI} & \textbf{CMI} & \textbf{SMI} &  \cellcolor{white}\\ \hline

- & Cer$_{def}$ & - & - & - & - & \multirow{4}{*}{\rotatebox{90}{\mbox{legacy}}}\\
\cellcolor{light_green} - & \cellcolor{light_green} Cer & \cellcolor{light_green} \ding{51} & \cellcolor{light_green} - & \cellcolor{light_green} - & \cellcolor{light_green} - & \\
Cer & Cer$_{def}$ & - & - & - & - & \\
\cellcolor{light_green} Cer & \cellcolor{light_green} Cer & \cellcolor{light_green} \ding{51} & \cellcolor{light_green} - & \cellcolor{light_green} - & \cellcolor{light_green} - & \\

\hline

- & DID$_{def}$ & - & - & \ding{51}/- & - &\\
\rowcolor{light_green} - & DID & \ding{51} & - & - & - & \cellcolor{white} \\
\multirow{2}{*}{DID} & \multirow{2}{*}{DID$_{def}$} & - & - & \ding{51} & \ding{51} & \cellcolor{white}\\
 &  & - & \ding{51} & - & - & \\
\rowcolor{light_green} & & \ding{51} & - & \ding{51} & \ding{51} & \cellcolor{white} \\
\rowcolor{light_green} \multirow{-2}{*}{DID} & \multirow{-2}{*}{DID} & \ding{51} & \ding{51} & - & - & \cellcolor{white}\\
- & DID$_{def}$+VC & - & - & \ding{51} & \ding{51} & \\
\rowcolor{light_green} - & DID+VC & \ding{51} & - & \ding{51} & \ding{51} & \cellcolor{white}\\
DID+VC & DID$_{def}$+VC & - & \ding{51}/- & \ding{51} & \ding{51} & \\
\rowcolor{light_green}DID+VC & DID+VC & \ding{51}  & \ding{51}/- & \ding{51} & \ding{51} & \cellcolor{white} \\
DID+VC & DID$_{def}$ & - & \ding{51}/- & \ding{51}/- & \ding{51}/- & \\
\rowcolor{light_green}DID+VC  & DID & \ding{51} & \ding{51}/- & \ding{51}/- & \ding{51}/- & \cellcolor{white}  \multirow{-12}{*}{\rotatebox{90}{DID/VC-enabled}} \\

\hline

 & & - & \ding{51} & - & - & \\
\multirow{-2}{*}{DID} & \multirow{-2}{*}{Cer$_{def}$} & - & - & \ding{51} & \ding{51} & \\
\rowcolor{light_green} DID & Cer & \ding{51} & - & \ding{51}/- & - & \cellcolor{white} \\ 
Cer & DID$_{def}$ & - & - & \ding{51}/- & - & \\ 
\rowcolor{light_green}Cer & DID & \ding{51} & - & - & - & \cellcolor{white} \\ 
DID+VC & Cer$_{def}$ & - & \ding{51}/- & \ding{51} & \ding{51} & \\ 
\rowcolor{light_green}DID+VC & Cer & \ding{51} & \ding{51}/- & \ding{51} & \ding{51} & \cellcolor{white} \\
Cer & DID$_{def}$+VC & - & - & \ding{51} & \ding{51} & \\
\rowcolor{light_green}Cer & DID+VC & \ding{51} & - & \ding{51} & \ding{51} & \cellcolor{white} \multirow{-9}{*}{\rotatebox{90}{hybrid}}\\\hline
\end{tabular}
\caption{Authentication scenarios and extensions}
\label{table:extensions}
\end{table}

During the VP validation, the DID docs of the VC issuers are resolved with the distributed ledger to verify the VCs' digital signatures. If the issuer's DID doc is already present in a local cache and still valid according to local security policies, a resolution is not required. VCs can expire if equipped with a validity period and/or can alternatively be revoked by the issuer. In case revocation is possible, the verifying entity has to check the revocation status of the VP during the VP validation. An Online Certificate Status Protocol (OCSP) or a similar mechanism does not exist for VPs yet. The holder proof of the VP, which materializes usually as a digital signature for the VP, can be kept empty if and only if the subject of the VC matches the DID that was used for the pseudo-anonymous authentication during the TLS handshake. In other words, a holder proof within a VP is not needed if the DID ownership was already proven during the TLS handshake. But there exist scenarios, e.g., where proxies are involved, in which the DID used for the pseudo-anonymous authentication can differ from the DID used as the subject of the VCs. In those cases, the holder proof is not empty and needs to be verified. It may even be necessary to obtain the DID doc of the VC's subject during the VP validation step if the DID is unknown. 

With DID Link, the entities are free to choose a suitable form of authentication based on the scenario-specific requirements, the presence of CA-issued X.509 certificates and/or DIDs/VCs, and potentially prior knowledge about each other. DID Link supports both, one-way and mutual authentication. It gives entities the opportunity to fall back on CA-issued X.509 certificates for the authentication in case self-signed X.509 certificates with DIDs are not supported. An entity can either authenticate with a CA-issued X.509 certificate, a self-issued and DID-based X.509 certificate only, or in combination with a VC. In hybrid mode, one party can make use of a self-issued and DID-based X.509 certificate or in combination with a VC to authenticate itself while the other party falls back on a CA-issued X.509 certificate. If the client does not demand the server to use a specific identifier for the authentication (by using the SNI extension), the server can provide a default CA-based X.509 certificate (Cer$_{def}$) or a default DID (DID$_{def}$) within a self-issued X.509 certificate. Table \ref{table:extensions} lists the numerous authentication options the entities can choose from together with the extensions present in the ClientHello and ServerHello messages. Since the extensions CPP and SPA consistently co-occur whenever at least one party expresses an intention to showcase VCs, they have been excluded from the table for the sake of readability.




\section{Evaluation}


DID Link represents an alternative authentication scheme for TLS 1.3, bridging the gap from authentication with centrally-approved to ledger-approved identification material. The practicability of DID Link depends upon how it performs in terms of speed against the conventional way of authentication in TLS with CA-issued X.509 certificates. However, in contrast to the legacy authentication scheme, DID Link comes in two flavors: a pseudo-anonymous authentication with DIDs only or an identification with VCs in addition. A meaningful assessment of its performance is therefore only possible by comparing both variants independently with their contemporary counterparts. The practical usefulness of DID Link also hinges on the bandwidth utilization of the resulting TLS channel. Hence, a comparison against an alternative data transport solution that also employs DIDs for authentication purposes is required to properly evaluate the proposed solution in terms of efficiency.



\subsection{Methodology}

From a conceptual point of view, a self-issued X.509 certificate with a subject of choice in the subject field is the closest analog to a self-issued x.509 certificate with a DID in the ASN field. In both cases, the certificate owner can prove ownership of the contained public key. Both types of certificates lack a proof of the binding's validity by a 3rd party. However, with VDR-anchored DIDs, this proof is provided by the VDR. With legacy self-issued X.509 certificates, there is no such standardized and trustful out-of-bound option to check the binding's validity. Hence, its validity can't be tested unless the identifier results from the key itself, e.g., through hashing as applied in the Bitcoin network to derive the wallet address from the public key. A derived identifier is of pseudo-anonymous nature like a DID. It does not share the property of being resolvable but its uniqueness is guaranteed probabilistically in the same way as with DIDs. The pseudo-anonymous authentication scheme with DIDs in DID Link was therefore evaluated by testing its performance against the authentication by means of a self-issued X.509 certificate with a derived identifier. If the pseudo-anonymous authentication is supplemented by an identification with VCs afterwards then DID Link accomplishes the same goal as an authentication with legacy X.509 certificates issued by CAs, namely proving to own a 3rd-party issued identifier. First, the DID ownership is proven during the TLS handshake with self-issued and DID-equipped X.509 certificates. Afterwards, the subject referred to by the DID is revealed with a DID-bound VC that contains the 3rd party attested identifier and optionally additional identity claims of the subject. The combination of a pseudo-anonymous authentication and an identification with DID Link is therefore evaluated by comparing it as a whole against the conventional authentication with CA-issued X.509 certificates. 

\begin{figure*}[t!]
\centerline{\includegraphics[scale=0.28]{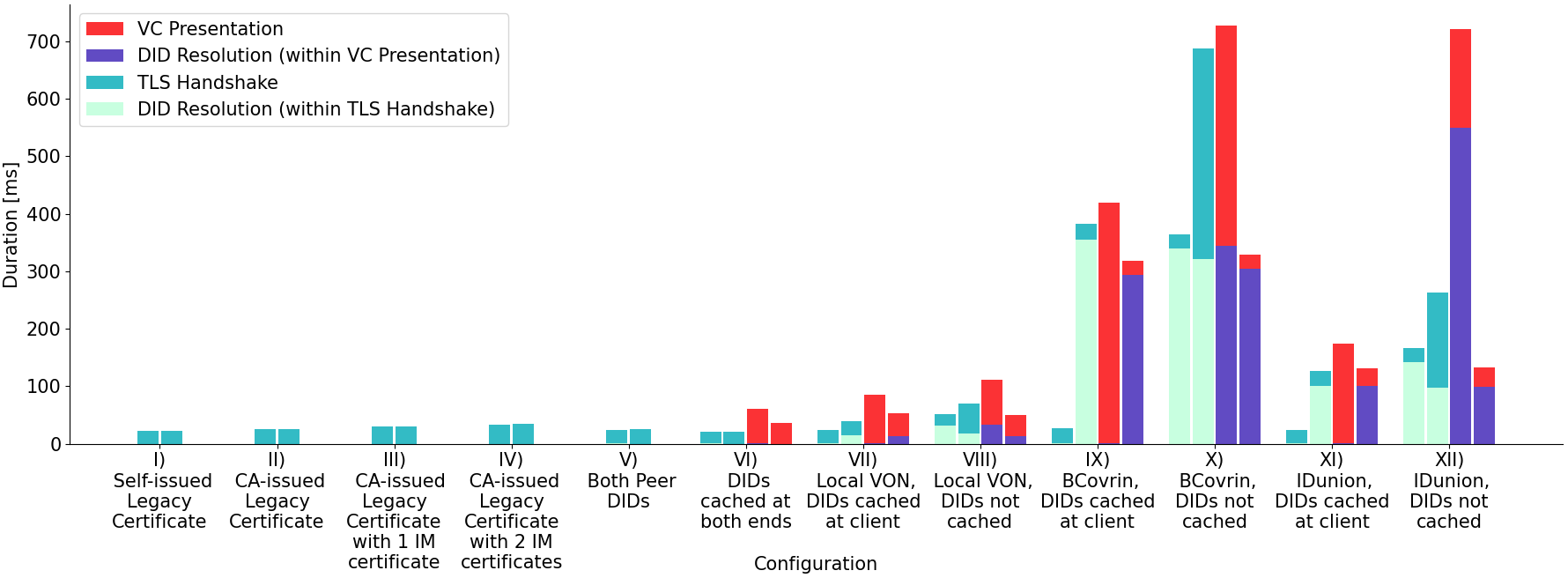}}
\caption{Average durations of the TLS handshakes and VC exchanges depending on the configuration after 1k runs for each configuration. The first (and third) column of each configuration shows the durations measured at the client while the second (and fourth) column the ones experienced at the server.}
\label{fig:identification_res}
\end{figure*}

For the evaluation of the bandwidth utilization, a TLS channel established with DID Link was compared against a communication link that uses DIDComm \cite{DIDComm:2020} instead. DIDComm is an application layer transport protocol built atop the concept of DIDs. It enables two parties to exchange messages in form of JSON Web Encryption (JWE) envelopes that are encrypted using the verification material found within each others' DID docs, ensuring end-to-end confidentiality, integrity, and authenticity. DIDComm operates without the need for established connections and is agnostic to the transport layer protocol, with implementations available across various programming languages. Despite being connectionless and not being a transport layer protocol, it was selected for the comparison due to its widespread use, the ongoing standardization by the DIF, and the lack of an available DID-enabled transport layer protocol. 


\subsection{Setup}
The proof of concept, comprising a client and a server, was implemented with Python 3.10 using OpenSSL 3.0.2 wrapped with pyOpenSSL 23.3. The VC presentation protocol was built from scratch, in compliance with the DIF Presentation Exchange 2.X.X specification. The VCs were encoded as SD-JWTs because of its simplicity compared to JSON-LD and other formats. For the performance comparison, the prototype relies on Veramo's implementation of DIDComm v2\footnote{\url{https://veramo.io/}}. The distributed ledgers used for the experiments were powered by Hyperledger Indy\footnote{\url{https://www.hyperledger.org/projects/hyperledger-indy}}. During the experiments, DID Link was tested with different ledger realizations. A locally deployed ledger based on the VON Network project\footnote{\url{https://github.com/bcgov/von-network}} was used to simulate a setup in which the client and server are each assumed to be co-located with a node of the ledger, making a ledger access practically a local rather than a remote operation. To simulate remote ledger access, two publicly accessible ledgers were used: the test ledger of the BCovrin Test Indy Network operated by the Province of British Columbia\footnote{\url{http://test.bcovrin.vonx.io/}} and the test ledger of the German IDunion project\footnote{\url{https://idunion.org/}}. The latter is commonly governed and operated by the IDunion project members, coming from various economic sectors such finance, transportation, telecommunication, and manufacturing. In the experimental setup, the client was operated on a machine with an Intel Core i5-6200U CPU with 2.30 GHz and 8 GB RAM while the server resided on a machine with an Intel Core i7-4790K CPU with 4.00 GHz and 32 GB RAM. Ubuntu-based KDE Neon 5.27 served as the operating system on both machines. They were located 7 km apart with an average ping time of 12.61ms and a bandwidth of 52 Mb/s (download) and 11 Mb/s (upload) of the client and respectively 57 Mb/s (download) and 22 Mb/s (upload) of the server. Access to the Internet was provisioned by Wi-Fi access points on both sites.

\subsection{Results}

In a first experiment, the average durations of the TLS handshakes for a mutual pseudo-anonymous authentication with legacy self-issued and with DID-based self-issued X.509 certificates were measured. The resulting average durations of the TLS handshakes for 1k runs per configuration, broken down by client and server are shown in Fig. \ref{fig:identification_res}. Config. I depicts the durations for a self-issued X.509 certificate with its derived identifier in the subject field. Config. VI to XII depict in dark green the total durations of the TLS handshakes with DID-based self-issued X.509 certificates. The proportion of time required for resolving a DID doc within the TLS handshake is illustrated proportionally in dark green. It is assumed that typically a client is well aware of the server's DID before contacting the server. In these cases, an up-to-date version of the server's DID doc is assumed to be pre-cached at the client, making a DID resolution during the TLS handshake not necessary. In configs. VI, VII, IX, and XI the DID docs of the server are assumed to be cached at the client while the client's DID doc is assumed to be cached at the server in config. VI. Config. V depicts a configuration in which both ends use Peer DIDs. Peer DIDs are special types of DIDs where the DID doc is encoded within the DID and is not necessarily anchored in a VDR \cite{DIF.peer}. The DID doc can therefore be resolved by decoding it from the DID itself. Peer DIDs are in particular useful when DIDs are ephemeral and the overhead of anchoring and resolving the DID doc from the VDR outweighs the unique benefit of a VDR-anchored DID doc to be modifiable, e.g., permitting to add, remove, or rotate verification material over time to improve security. 

The results show that if VDR-anchored DIDs are cached at the client and server (config. V) or peer DIDs are used instead (config. VI), the TLS handshake with DID Link performs similar in terms of speed as their legacy counterpart (config. I). As soon as a DID doc resolution via a VDR is required, the TLS handshake with DID Link takes significantly longer. For example, a resolution via a locally deployed VON ledger at the server in config. VIII increases the total handshake time in comparison to config. VI by a factor of 3.23 while it increases by factor of 31.58 with the BCoverin ledger in config. X and 12.12 with the IDunion ledger in config. XII. The reasons for the high resolution times via a remote ledger could be manifold, ranging from the implementation of the resolution logic and the ledger itself, over the communication latency with the ledger and the amount of transactions in the ledger, to the genesis file that specifies the settings for the ledger access. The prototype was operated in Germany, with a low latency towards the Germany-based IDunion ledger and a higher latency towards the BCoverin ledger operated in western Canada. The BCoverin ledger persisted approx. 640k transactions while the IDunion test ledger contained only approx. 117k transactions. During the evaluation, the Indy CLI driver\footnote{\url{https://github.com/hyperledger/indy-cli-rs}} provided by the Hyperledger Indy project was used to resolve the DID docs as it turned out in prior experiments with a locally deployed VON ledger that the resolution via a locally deployed Universal Resolver\footnote{\url{https://dev.uniresolver.io/}} or directly with the Indy driver of the Universal Resolver is on average 2.03 respectively 1.89 times slower than a resolution with the Indy CLI. A genesis file points to a set of ledger nodes usable for the resolution. This selection is of utmost importance since slow or non-reachable ledger nodes in the genesis file negatively impact the overall resolution performance, as experienced with different genesis file versions for the IDunion ledger. 


In a second experiment, the durations of the mutual identification step after the TLS handshake were measured in addition for the configs. VI to XII. Outliers (approx. 1\%) were removed since non-reachable ledger nodes during the DID resolution led to timeouts that significantly skewed the average durations. Since the pseudo-anonymous authentication together with the identification step of DID Link reassembles an authentication with a CA-issued X.509 certificate, the durations for the latter were measured in configs. II to IV for comparison reasons. The messages exchanged by the presentation protocol during the identification in the proof of concept differs slightly to the sequence of messages sketched schematically in Fig. \ref{fig:concept}. The DIF Presentation Exchange specification demands that a message containing the VP is the response to a verifier's request to present a VP. Hence, in the prototype, each party triggers the other  party's presentation of a VP by requesting it explicitly. However, the mutual VP presentations in the prototype were conducted in parallel to improve the identification step's performance.

\begin{figure}[t!]
\centerline{\includegraphics[scale=0.295]{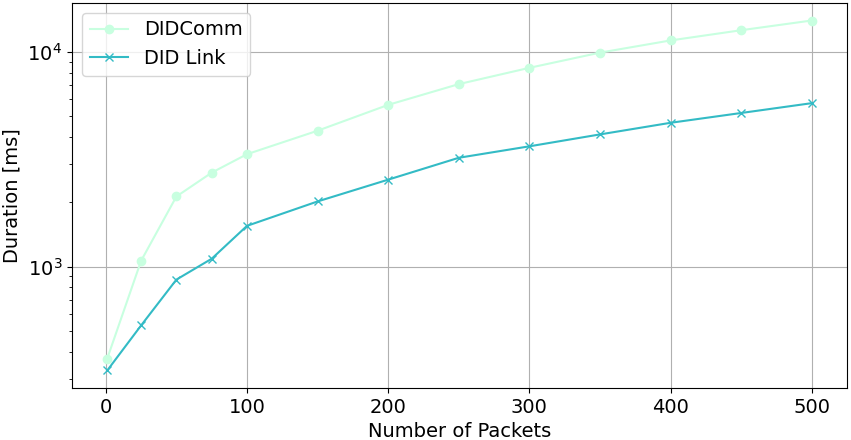}}
\caption{Average time to send a bulk of packets with a packet payload size of 10 kB within the same session.}
\label{fig:performance}
\end{figure}


The results show that when compared with an authentication based on CA-issued X.509 certificates, it takes at the client about 3.19 and 6.43 times longer to identify a subject with DID Link in configs. VI and VII and about 42.94 and 34.86 times longer if a remote ledger is queried for the DID docs in configs. X and XII. The times to resolve the DID docs of the client, server, and the issuers of the VCs dominate the overall time of the pseudo-anonymous authentication and identification steps of DID Link, making the DID resolution the major bottleneck of the proposed solution. In all configs. from VI to XII, it seems as it takes significantly longer for the client to identify the server than vice versa. This phenomenon results from the fact that the client has to wait for the server to process the client's Certificate, CertificateVerify, and Finished messages before the server can proceed with the identification step. Since the client emits the first identification message, namely the VC presentation request, immediately after emitting the Finished message, it is already engaged in the identification while the server is still busy in the TLS handshake. However, an identification step with VCs is assumed to happen only once two entities have to become known to each other. Once this is done, a pseudo-anonymous authentication with DIDs is sufficient since the subjects the DIDs refer to are known. If commonly agreed upon DID doc update policies are established upon all participants of a ledger, DID docs might only change at regular or defined points in time so that most of the time locally cached versions of DID docs are sufficient for a secure TLS connection establishment. A pseudo-anonymous authentication in config. VI with DID Link is by a factor of 1.05 even slightly faster compared to the authentication in config. I. In contrast to DID Link and due to tight coupling of a CA-issued identifier and the corresponding key material in a signed artifact, the authentication with CA-issued X.509 certificates needs to be repeated at every TLS connection establishment.  
 






\setlength{\tabcolsep}{0.62em} 
{\renewcommand{\arraystretch}{1.0}
\begin{table}[t]
\vspace{0.2cm}
\centering
\begin{tabular}{|c|c|c|c|c|c|c|}
\hline
\rowcolor{strong_green} \textbf{payload} &  & \multicolumn{5}{|c|}{\textbf{\# of packets}} \\ 

\rowcolor{strong_green}  \textbf{[kB]} & \multirow{-2}{*}{\textbf{variant}} & \textbf{1} & \textbf{2} & \textbf{3} & \textbf{4} & \textbf{5}\\

\hline
 & $\Diamond$ & 4.479 & 4.506 & 4.533 & 4.560 & 4.587 \\
\multirow{-2}{*}{0.001} & $\blacklozenge$ & 1.392 & 2.336 & 3.220 & 4.134 & 5.048 \\

\rowcolor{light_green}  & $\Diamond$ & 4.488 & 4.524 & 4.560 & 4.596 & 4.632 \\
\rowcolor{light_green} \multirow{-2}{*}{0.01}  & $\blacklozenge$ & 1.404 & 2.330 & 3.256 & 4.182 & 5.108 \\

 & $\Diamond$& 4.578 & 4.704 & 4.830 & 4.956 & 5.082 \\
\multirow{-2}{*}{0.1} & $\blacklozenge$ & 1.524 & 2.570 & 3.616 & 4.662 & 5.708 \\

\rowcolor{light_green} & $\Diamond$ & 5.478 & 6.504 & 7.662 & 8.622 & 9.714 \\
\rowcolor{light_green}\multirow{-2}{*}{1} & $\blacklozenge$ & 2.856 & 5.234 & 7.612 & 9.990 & 12.368 \\

 & $\Diamond$ & 14.808 & 25.560 & 36.246 & 46.998 & 57.090 \\
\multirow{-2}{*}{10} & $\blacklozenge$ & 15.912 & 31.238 & 45.868 & 60.036 & 76.702\\
\hline
\end{tabular}
\caption{Kilobytes transferred with DID Link ($\Diamond$) or DIDComm ($\blacklozenge$) depending on number of packets per session and payload size}
\label{table:didcomm}
\end{table}

Finally, the performances of TLS 1.3 (with DID Link) was compared with DIDComm version 2 in terms of bandwidth utilization and the average time to transfer data in the experimental setup. For a fair comparison, DIDComm applied TCP as the transport layer protocol. The total bytes transferred by the client towards the server were measured for different payload sizes and number of packets per session. It includes two session tickets for TLS with DID Link of total size 1.5 kB for a potential session resumption. The results are shown in Table \ref{table:didcomm}. For the data transfer time measurements, the client established a TLS connection (with DID Link) to the server and transferred a certain amount of packets via the same session. A TCP session was established instead of TLS for the measurements with the DIDComm protocol. The average time for the client to send a bulk of packets to the server in the experimental setup depending on the amount of packets per bulk after 50 runs are shown in Fig. \ref{fig:performance}. 

For payload sizes up until 1 kB, the initial protocol overhead with DID Link for the transfer of a single packet is significantly higher than the one experienced with DIDComm because of the required TLS handshake. However, once the session is established, the protocol overhead with DID Link/ TLS becomes smaller in comparison to DIDComm. The point at which a connection with DIDConnect utilizes the channel better depends on the payload size. It needs 5 packets with payload size of 1 to 100 bytes for a TLS connection to be more efficient than DIDComm or 4 packets for payload size 1 kB. Since DIDComm lacks a concept of a session there exists no agreed upon symmetric key for the ciphering and deciphering of all packets within a session. Instead, DIDComm v2 applies the Elliptic-curve Diffie–Hellman One-Pass Unified Model (ECDH-1PU) key derivation process per message, making it more computational expensive to handle ciphered communication than using a symmetric session key for all messages of a session like in TLS. The comparably higher costs in terms of overhead and CPU usage lead to significantly higher data transfer times for DIDComm in comparison to DID Link/TLS, especially in case a high amount of packets is transferred per session. In this TCP-based experimental setup with a reliable communication link, a TLS channel established with DID Link therefore enables a quicker end-to-end data transfer after only few packets for payload size below 1 kB and instantly for payload
sizes above 10 kB  compared to a message transfer based on DIDComm. For example, 100 packets with payload size of 10 kB each is transferred 2.16 times faster via a DID Link-established TLS channel than via DIDComm over a TCP channel.




\section{Conclusion}
This article presented DID Link, a novel authentication scheme for TLS 1.3 that comes in two flavors, namely a basic version for the pseudo-anonymous authentication with DIDs and an advanced version for the identification with VCs. DID Link extends TLS 1.3 with the option to let entities authenticate alternatively with artifacts of a decentrally managed digital identity, all without the need to fundamentally redesign the well approved authentication procedure of TLS. It introduces a new identification sub layer and custom extensions to the initial TLS handshake messages in the frame of permissible modifications. The identification sub layer delineates the identification-specific communication from the application data, while the custom extensions give both parties early on the opportunity to agree on common DID- and VC-specific parameters for the authentication. The adherence to the TLS 1.3 standard ensures that DID Link can be seamlessly integrated into existing legacy systems while  
benefiting from a certificate-based authentication procedure that is widely used, well established and in particular thoughtfully investigated to ensure security across diverse platforms and applications. DID Link facilitates the decentralization of the Web by enabling entities to autonomously and in a self-sovereign manner manage their identifiers and identity claims and use them for the authentication at the transport layer with TLS, without having to sacrifice the fundamental security features TLS has to offer, such as confidentiality, integrity, and authenticity. 



Nevertheless, entrusting a commonly operated and governed distributed ledger with one of the essential functions of a centralized CA comes with costs. The out-of-band verification of the identifier and key binding via the ledger leads to a significant prolongation of the authentication process. But once the identification is done, the entities can fall back on the pseudo-anonymous authentication with cached DID docs, which performs even slightly better than the authentication with CA-issued X.509 certificates. Moreover, a TLS channel established with DID Link outperforms a DIDComm link in terms of speed by a factor of more than 2 at latest after a handful of messages has been transmitted. As a next step, the VP used to carry identity claims can be equipped additionally with VCs for the authorization such like an access token, making it possible to unify and harmonize access control procedures within a single step on the transport layer. However, the exchange of VCs for access control purposes with DID Link requires a further security analysis, as does the sharing of verification material via a common and cross-domain system in form of a distributed ledger.

\bibliographystyle{IEEEtran}  
\bibliography{references}  

\end{document}